\documentclass{pramana}

\usepackage{graphicx,amsmath,bm}
\usepackage[font=scriptsize]{caption}
\usepackage{subcaption}
\usepackage{amsfonts,amssymb,amsmath}
\usepackage{graphicx}
\usepackage{mathtools, cuted}
\usepackage{lipsum, color}
\usepackage{flushend}
\begin{document}

%%paper title
%%For line breaks \\ can be used within title
\title{Quadrature operator eigenstates and wavefunctions of \\ $f$-deformed  oscillators}

%%author names are separated by comma (,)
%%use \and before the last author name
%%\textsuperscript{number} is used for affiliation
%%use a * along with the number separated by comma
%% for the  author for correspondence

\author{S ANUPAMA\textsuperscript{1}, ADITI PRADEEP\textsuperscript{2}, ADIPTA PAL\textsuperscript{3}\and C SUDHEESH\textsuperscript{1,*}}
\affilOne{\textsuperscript{1} Department of Physics, Indian Institute of Space Science and Technology,
Thiruvananthapuram, 695 547, India\\}
\affilTwo{\textsuperscript{2} Department of Physics and Astronomy, University of British Columbia, Vancouver, BC V6T 1Z1, Canada\\}
\affilThree{\textsuperscript{3} Department of Physics, Indian Institute of Science Education and Research,
Kolkata, 741 246, India}

%%escape two column mode for title, affiliation and abstract
%%by giving \twocolumn command as shown

\twocolumn[{

\maketitle

%%include \corres to print the corresponding author Email id
\corres{sudheesh@iist.ac.in}

%%include \msinfo for
%%manuscript information such as
%%received, revised and accepted dates
%%
%\msinfo{1 January 2015}{1 January 2015}{1 January 2015}

%%abstract
\begin{abstract}
This paper is dedicated to finding the quadrature operator eigenstates and wavefunctions of the most general $f$-deformed oscillators. A definition for quadrature operator for deformed algebra is derived to obtain the quadrature operator eigenstates. A new set of polynomials
 are obtained using this quadrature operator and these polynomials are used to find explicitly the wavefunctions of the deformed oscillators.
 We have plotted wavefunctions for three different  types of deformations and compared it with the wavefunctions of the non-deformed oscillator.  Our result will immensely help the research groups  working in the quantum state reconstruction and quantum information theory of deformed states.
\end{abstract}

%%insert keywords separated by comma using \keywords{words}

%%include \pacs{number} to print the PACS number
\pacs{ }

}]
%%close the twocolumn escape here

%%include \doinum{number}for the DOI number in the header
%%include \volnum{number} for the volume number in the header
%%include \year{yyyy} for  year of publication in the header
%%include \pgrange{num--num} page range of article in the header
%%include \artcitid{num} for the article citation id
%%include \lp to print last page of the article
%%include \setcounter{page}{pagenum} for the exact starting page of the article

%\doinum{\#-\#-\#-\#}
%\artcitid{\#\#\#\#}
%\volnum{\#}
%\year{\#}
%\pgrange{\#--\#}
%\setcounter{page}{\#}
%\lp{\#}

\section{Introduction}
The generalized deformed $f$-oscillators with deformation function  and deformed boson algebra  have been making their appearance in the description of many physical phenomena recently.  Special cases of the $f$- deformation, like the $q$-deformation have been used recently to study the cosmic microwave background radiation \cite{zeng}. $q$-deformed bosons have been explored as tools for the realization of quasibosons, and modelling quasiparticles with potential applications in subnuclear physics and quantum information theory \cite{gavrilik}. The energy density distribution of bosons obeying the $q$-deformations of the harmonic oscillator algebra have been studied in \cite{martin}. Quantum logic gates have been constructed using $q$-deformed harmonic oscillator algebras \cite{altinas}. ($p,q$)-deformed bosons, another class of $f$-deformed states, have been used to study deformed Bose gases with critical temperature ratio $T_c^{p,q}/T_c$ which explicitly depends on the deformation parameters $p$ and $q$ \cite{gavrilik2}. The solutions of deformed Einstein equations and quantum black holes which use $q$ and ($p,q$)-deformed algebras have been obtained recently \cite{dil}. Low temperature behaviour of deformed fermion gas models have been used to find interactions of quasiparticles which are having applications in nanomaterials \cite{algin}. It has been shown that $q$-deformed bosonic exciton gas constitutes the high density limit of Frenkel excitons which may provide valuable insight into Frenkel excitons inside nanomaterials \cite{zeng_cheng}. A class of Fibonacci oscillators which uses deformed algebras have found applications in the Debye model to study thermodynamics of crystalline solids \cite{marinho}  and the Landau diamagnetism \cite{marinho2}. Also, a recent study of the behaviour of math type $q$-deformed harmonic oscillator clearly shows the signature of chaos in the system \cite{Aditi}. The  $f$-deformed Hamiltonian has been studied for the anti-Jaynes-Cummings  model, with applications in quantum optic interferometers \cite{setare}. Another recent research shows that the $f$-deformed Dirac oscillator can describe the electrons in a non-linear zig- zag graphene nanoribbon and it causes significant difference between Landau levels in the $f$-deformed oscillator and the non-deformed one \cite{setare2}. A mathematical procedure to obtain analytical expressions for a general class of $q$-deformed coherent states associated with the different patterns of the energy spectrum
exhibited by the nonlinear $f$-oscillator was presented in the literature \cite{marcelo} and deformed photon-added nonlinear coherent states were constructed using nonlinear coherent states \cite{roman}.
\par We observe that although the general deformed $f$-oscillators with their different classes have generated much importance in recent years, their wavefunctions, so far remain unknown. A notable attempt has been made to derive a generalised wavefunction which encompasses the wavefunctions for the Macfarlane and Dubna type oscillators \cite{sogami}. However, this expression is limited by its intricate and restricted access to the parameters involved, thus being unable to provide a clear-cut wavefunction to study the system. The analytical wavefunction of a non-deformed harmonic oscillator can be easily evaluated from the knowledge of the deformed Hermite polynomials. Unfortunately, an explicit expression for the deformed Hermite polynomials is unknown when subject to $f$-deformation but some attempts have been made on this front over the years. In another work\cite{lorek}, the authors have tried to approach the problem by evaluating the analytical form of the $q$-deformed Hermite polynomial in terms of the position operator $\hat{X}$. This result is again limited by the lack of knowledge of the nature and form of the operator $\hat{X}$. However, recently, using the expression for the $q$-deformed quadrature operator $\hat{X}_{\theta}$, the expression for $q$-deformed Hermite polynomial has been found \cite{jayakrishnan}.  The  paper \cite{jayakrishnan} deals only with the $q$-deformed oscillator but not the general $f$-deformed case and does not discuss in detail the wavefunctions of the  oscillator. %In \cite{16} the properties of these deformed Hermite polynomials have been described.
The quadrature wavefunctions  of quantum systems are very important and essential  for the better understanding  of physical systems  and reconstruction of quantum systems from experiments. In this paper, we have defined the $f$-deformed quadrature operator and obtained the quadrature wavefunctions for different classes of deformed oscillators with different degrees of deformation.
\par
This paper is organised in the following way.  A review of $f$-oscillators is included in Section 2. Section 3 starts with the definition of quadrature operator and the derivation of  eigenstates of the quadrature operator.  Recurrence relations for  new polynomials are derived in this Section. We end this section with an analytical derivation of    $f$-oscillator wavefunctions and their  graphical representations.  We summarise our results  in section 4.

\section{Review of $f$-oscillators}
We  recall that  the algebra associated with the quantum mechanical harmonic oscillator is defined by the canonical  commutation relations
\begin{equation}
\label{lie}
[\hat{a},\hat{a}^\dagger]=\hat{a}\hat{a}^\dagger-\hat{a}^\dagger \hat{a}=\hat{I},\qquad
[\hat{a},\hat{I}]=[\hat{a}^\dagger,\hat{I}]=0,
\end{equation}
where $\hat{I}$ is the identity operator and \(\hat{a}\) and \(\hat{a}^\dagger\) are respectively the lowering and raising operators for the oscillator energy eigenstates. The following properties hold for these operators:\\
\begin{eqnarray*}
\hat{a}|n\rangle&=&\sqrt{n}|n-1\rangle, \quad {\tt for} \; n=1,2,3,...,\\
\hat{a}^\dagger|n\rangle&=&\sqrt{n+1}|n+1\rangle, \quad {\tt for} \;n=0,1,2,3... ,
\end{eqnarray*}
with $\hat{a}|0\rangle=\Omega$, where $\Omega$ is the null vector.
The number operator $\hat{n}=\hat{a}^\dagger \hat{a}$ satisfies the following commutation relations:
$$[\hat{n},\hat{a}]=-\hat{a},\qquad\qquad[\hat{n},\hat{a}^\dagger]=\hat{a}^\dagger.$$\par
The authors of \cite{mizrahi} have defined the operator $\hat{A}$ and its adjoint $\hat{A}^\dagger$ as a non-linear expansion of the usual harmonic oscillator operators $\hat{a}$ and $\hat{a}^\dagger$:\\
\begin{equation}
\hat{A}=\hat{a}f(\hat{n}),\qquad\qquad \hat{A}^\dagger=f^\dagger(\hat{n})\hat{a}^\dagger.\qquad
\end{equation}

\par The operators $\hat{A}$ and $\hat{A}^{\dagger}$ in terms of a function $f(\hat{n})$ define a new class of oscillators called as $f$-oscillators which reduces to the usual harmonic oscillator operators $\hat{a}$ and $\hat{a}^{\dagger}$ when $f(\hat{n})=\hat{I}$. These new operators satisfy the following commutation relation \cite{manko}:\\
\begin{equation}
[\hat{A},\hat{A}^\dagger]=\phi(\hat{n}),
\end{equation}
where $\phi(\hat{n})=f(\hat{n}+1)f^{\dagger}(\hat{n}+1) (\hat{n}+1)-f(\hat{n})f^{\dagger}(\hat{n}) \hat{n}$. They also obey the $Q$-commutation relation
\begin{eqnarray}
[\hat{A},\hat{A}^\dagger]_Q=\hat{A}\hat{A}^\dagger-Q\hat{A}^\dagger \hat{A}\quad \quad \nonumber\\
=f(\hat{n}+1)f^{\dagger}(\hat{n}+1) (\hat{n}+1)-Qf(\hat{n})f^{\dagger}(\hat{n})\hat{n},
\label{eq_Qcomm}
\end{eqnarray}
where $Q$ is a general deformation parameter and different forms of $Q$ provide different types of deformation; we will be considering a few specific types of deformations in this paper. The above relation reduces to the  canonical commutation relation (\ref{lie}) when $Q\longrightarrow1$ and $f(\hat{n})=I$.
\par
We can now define a new set of eigenstates $|n\rangle_f$ that form a complete orthonormal basis in the $f$-deformed Fock space provided that there exists a deformed number operator $\hat{N}=\hat{A}^{\dagger}\hat{A}$ with eigenvalue $[n]$ having the form
\begin{equation}
[n]=|f(n)|^2n,
\end{equation}
such that,
\begin{equation}
\hat{N}\,|n\rangle_f=\hat{A}^\dagger \hat{A}\,|n\rangle _f=[n]|n\rangle _f,
\qquad _f\langle m|n\rangle _f=\delta_{mn}.
\end{equation}
The action of $\hat{A}$ and $\hat{A}^{\dagger}$ on the deformed Fock states $|n\rangle_f$ are given by
\begin{equation}
\begin{split}
\hat{A}\,|n\rangle_f=\sqrt{[n]}\,|n-1\rangle_f, \quad \hat{A}\,|0\rangle_f=0 \;{\tt and}\\
\hat{A}^\dagger\,|n\rangle_f=\sqrt{[n+1]}\,|n+1\rangle_f.
\end{split}
\end{equation}\label{oper_eq}

\par Now we review some special cases of $f$-oscillators which we will be considering in our study:
\begin{itemize}
\item The math-type $q$-deformation obeys the commutation relation \cite{arik}:
\begin{equation}\label{mathq_comm}
\hat{A}\hat{A}^\dagger-q^2\hat{A}^\dagger \hat{A}=I,
\end{equation}
where, $0<q<1$ with $Q=q^2$. Here,
\begin{equation}\label{fn_q2}
[n]=\frac{1-q^{2n}}{1-q^2}.
\end{equation}
\item The $(p,q)$-deformation follows the commutation relation \cite{chakrabarti}:
\begin{equation}\label{pq_comm}
\hat{A}\hat{A}^\dagger-p\hat{A}^\dagger \hat{A}=q^{-\hat{n}}.
\end{equation}
Here,  $Q=p$ and
\begin{equation}\label{fn_pq}ref
|f(n)|^2 n=[n]=\frac{q}{1-pq}(q^{-n}-p^{n}).
\end{equation}
\item The physics-type $q$-deformation \cite{biedenharn} is actually a special case of (\ref{pq_comm}) with $p=q$, giving us
\begin{equation}\label{phyq_comm}
\hat{A}\hat{A}^\dagger-q\hat{A}^\dagger \hat{A}=q^{-\hat{n}},
\end{equation}
where $q>1$. Here, $Q=q$ and  putting $p=q$ in (\ref{fn_pq}), we get
\begin{equation}\label{fn_phyq}
|f(n)|^2 n=[n]=\frac{q}{1-q^2}(q^{-n}-q^{n}).
\end{equation}
\end{itemize}
Similar relations can be obtained for other maths-type \cite{quesne,brezi} and physics-type \cite{macfarlane} deformed oscillators.

\section{The $f$-deformed quadrature operator and wavefunctions of the $f$-deformed oscillator}
Consider the position operator  $X$ and  momentum operator $P$ for the $f$-oscillator in terms of the $f$-deformed ladder operators \cite{bijan,dey}
\begin{equation}
\hat{X}=\alpha(\hat{A}^\dagger+\hat{A}),\qquad
\hat{P}=i\beta(\hat{A}^\dagger-\hat{A}),
\label{x,p}
\end{equation}
where $\alpha, \beta \in \Re$. Rearranging (\ref{x,p}), we obtain the expressions for $\hat{A}$ and $\hat{A}^\dagger$ in terms of $\hat{X}$ and $\hat{P}$. Substituting these expressions for $\hat{A}$ and $\hat{A}^{\dagger}$ into (\ref{eq_Qcomm}), for $\alpha=\beta=\frac{\sqrt{1+Q}}{2}$, we obtain the following  commutation relation for the deformed operators $\hat{X}$ and $\hat{P}$:
%\begin{equation}
\begin{eqnarray}
[\hat{X},\hat{P}]=i[f(\hat{n}+1)f^{\dagger}(\hat{n}+1) (\hat{n}+1)-Qf(\hat{n})f^{\dagger}(\hat{n}) \hat{n}\nonumber\\-\frac{1-Q}{1+Q}(X^2+P^2)].
\end{eqnarray}

%\end{equation}
We can see that the above relation gives distinct  deformed algebras for the examples mentioned in the previous section when their respective functions $|f(n)|^2$ and respective deformation parameters $Q$ are substituted.

The homodyne quadrature operator  corresponds to the canonical commutation relation given in  (\ref{lie}) is \cite{vogel,yurke,yuen,yuen2}

\begin{equation}
\label{quad}
\hat{x}_\theta=\frac{1}{\sqrt{2}}(\hat{a}e^{-i\theta}+\hat{a}^\dagger e^{i\theta}).
\end{equation}
where $\theta$ is the phase of the local oscillator associated with the homodyne detection setup, such that $0\leq \theta$ $\leq 2\pi$. It is easy to see that we obtain the canonical position and momentum operators $\hat{x}$ and $\hat{p}$ when  $\theta=0$ and $\theta=\pi/2$, respectively.

It can be easily verified using the photon-number difference in the two output channels of the beam splitter in homodyne detection setup
that  the $f$-deformed homodyne quadrature operator takes the form:
\begin{equation}\label{xtheta}
\hat{X}_\theta=\frac{\sqrt{1+Q}}{2}(\hat{A}e^{-i\theta}+\hat{A}^\dagger e^{i\theta}),
\end{equation}
and it becomes deformed  position and momentum operators $\hat{X}$ and $\hat{P}$ when  $\theta=0$ and $\theta=\pi/2$, respectively.
In the limit, $f(\hat{n})=I$ and $Q\longrightarrow1$, the deformed quadrature operator $\hat{X}_\theta$ reduces to the non-deformed quadrature operator given in (\ref{quad}). Using the definition of the deformed quadrature given above, we find the quadrature wavefunction
of $f$-oscillator in the following part of the paper.

The eigenstate of the deformed quadrature operator is represented by $|X_\theta\rangle$ and it has an eigenvalue $X_\theta$ given by
\begin{equation}\label{eig_eq}
\hat{X}_\theta|X_\theta\rangle=X_\theta|X_\theta\rangle.
\end{equation}
These eigenstates can be represented in the deformed Fock state basis of the $f$-deformed oscillator as:
\begin{equation}\label{fock_rep_xtheta}
|X_\theta\rangle=\sum\limits_{n=0}^{\infty}|n\rangle_{f f}\langle n|X_\theta\rangle,
\end{equation}
where,
\begin{equation}\label{quad_Fock}
\langle X_\theta|n\rangle_f=\Psi_{n}(X_\theta)
\end{equation}
 is the wavefunction of the $f$-oscillator for a given deformed Fock state $|n\rangle_f$ in the quadrature representation. We obtain the excited state wavefunctions in terms of the ground state wavefunction in what follows. This simplifies the problem at hand by reducing the unknowns to just the ground state wavefunction.
\par The eigenvalue equation (\ref{eig_eq})  of the operator  $\hat{X}_\theta$ allows us to derive a recurrence relation obeyed by the wavefunctions $\Psi_{n}(X_\theta)$. Using (\ref{oper_eq}), (\ref{xtheta}) and (\ref{eig_eq}) we get,
\begin{eqnarray}%\label{exp_x}
%\begin{aligned}
_f\langle n|\hat{X}_\theta|X_\theta\rangle &=& X_\theta \overline{\Psi}_{n}(X_\theta)\nonumber\\
&=&\frac{\sqrt{1+Q}}{2} _f\langle n|(\hat{A}e^{-i\theta}+\hat{A}^\dagger e^{i\theta})|X_\theta \rangle \nonumber
%&=&\frac{\sqrt{1+Q}}{2}\left(\sqrt{[n+1]} \; e^{-i\theta} \; _{f}\langle n+1|X_\theta\rangle +%\sqrt{[n]} \; e^{i\theta} \; _{f}\langle n-1|X_\theta\rangle\right)\nonumber\\
%&=&\frac{\sqrt{1+Q}}{2}\left(\sqrt{[n+1]} \; e^{-i\theta} \; \overline{\Psi}_{{n+1}}(X_\theta)
%+\sqrt{[n]} \; e^{i\theta} \; \overline{\Psi}_{{n-1}}(X_\theta)\right),
%\end{aligned}
\end{eqnarray}
\\
On substitution, we get
\\
\begin{strip}
\begin{eqnarray}\label{exp_x}
_f\langle n|\hat{X}_\theta|X_\theta\rangle=\frac{\sqrt{1+Q}}{2}\left(\sqrt{[n+1]} \; e^{-i\theta} \;  \overline{\Psi}_{{n+1}}(X_\theta)+\sqrt{[n]} \; e^{i\theta} \; \overline{\Psi}_{{n-1}}(X_\theta)\right),
\end{eqnarray}
\end{strip}
for $n=1,2,3...$, and
\begin{equation}\label{Psi1conj}
X_\theta \; \overline{\Psi}_{0}(X_\theta)=\frac{\sqrt{1+Q}}{2}\Big(\sqrt{[1]} \; e^{-i\theta} \; \overline{\Psi}_{{1}}(X_\theta)\Big),
\end{equation}
 where $\overline{\Psi}_{{n}}(X_\theta)$ is the complex conjugate of the wavefunction corresponding to the deformed state $|n\rangle_f$.
The complex conjugate of (\ref{exp_x}) gives us a two term recurrence relation for $\Psi_{n+1}(X_\theta)$ in the quadrature representation:
\begin{eqnarray}\label{Psi_rec}
\Psi_{n+1}(X_\theta)=\frac{e^{-i\theta}}{\sqrt{[n+1]}}\bigg[\frac{2}{\sqrt{1+Q}}X_\theta \; \Psi_{n}(X_\theta)\nonumber\\
-\sqrt{[n]} \; \Psi_{n-1}(X_\theta) \; e^{-i\theta}\bigg]
\end{eqnarray}
for $n=1,2,3,...,$ with
\begin{equation}\label{Psi1}
\Psi_{1}(X_\theta)=\frac{e^{-i\theta}}{\sqrt{[1]}}\frac{2X_\theta}{\sqrt{1+Q}}\Psi_{0}(X_\theta).
\end{equation}
Using (\ref{Psi_rec}) and (\ref{Psi1}), we obtain some of the excited state  wavefunctions in the quadrature representation:
\begin{eqnarray}
\label{Psi2}
\Psi_{2}(X_\theta)=\frac{e^{-2i\theta}}{\sqrt{[2]}}\bigg[\frac{2X_\theta}{\sqrt{1+Q}}\bigg(\frac{2X_\theta}{\sqrt{[1](1+Q)}}\bigg)\nonumber\\
-\sqrt{[1]}\bigg]\Psi_{0}(X_\theta),
\end{eqnarray}
\begin{eqnarray}\label{Psi3}
\Psi_{3}(X_\theta)=\frac{e^{-3i\theta}}{\sqrt{[3]}}\bigg[\frac{2X_\theta}{\sqrt{1+Q}}\frac{1}{\sqrt{[2]}}\bigg(\frac{2X_\theta}{\sqrt{1+Q}}\nonumber\\
\frac{2X_\theta}{\sqrt{[1](1+Q)}}-\sqrt{[1]}\bigg)-\sqrt{[2]}\nonumber\\
\frac{2X_\theta}{\sqrt{[1](1+Q)}}\bigg]\Psi_{0}(X_\theta).
\end{eqnarray}
Similarly, we can obtain any other excited state using the recurrence relation. We observe that form of the solutions are similar to the solutions obtained for the $q$-deformed state \cite{jayakrishnan} with the following changes: $q^2$ has been replaced by a general deformation parameter $Q$ and $[n]$ has the value $|f(n)|^2n$.

Now, we can write  the analytical expression for the $f$-deformed Fock state $|n\rangle_f$ in the quadrature basis as
\begin{equation}\label{Psi_nf}
\Psi_{n}(X_\theta)=e^{-in\theta} \; J_{n}(X_\theta) \; \Psi_{0}(X_\theta),
\end{equation}
where $\Psi_{0}(X_\theta)$ is the ground state wavefunction and the new polynomial $J_{n_f}(X_\theta)$ is defined by the following recurrence relation:
\begin{equation}\label{J_nf}
J_{n+1}(X_\theta)=\frac{1}{\sqrt{[n+1]}}\bigg[\frac{2X_\theta}{\sqrt{1+Q}} \; J_{n}(X_\theta)-\sqrt{[n]} \; J_{n-1}(X_\theta)\bigg].
\end{equation}
It is very clear from the above expression that $J_{n}(X_\theta)$ is dependent on $Q$ and therefore, varies for each case that is being studied. The initial two terms of $J_{n}(X_\theta)$ are:
\begin{equation}
J_{0}(X_\theta)=1,
\end{equation}
\begin{equation}
J_{1}(X_\theta)=\frac{2X_{\theta}}{\sqrt{[1](1+Q)}}.
\end{equation}
Using  Favard's theorem, we are able to prove that the above recurrence relation gives birth to a class of orthogonal polynomials. From (\ref{J_nf}), we find the second and third terms of $J_{n}(X_\theta)$ to be
\begin{equation}
J_{2}(X_\theta)=\frac{4X_{\theta}^2-(1+Q)[1]}{\sqrt{[2]!(1+Q)^2}},
\end{equation}
\begin{equation}
J_{3}(X_{\theta})=\frac{8X_{\theta}^3-2(1+Q)([1]+[2])X_{\theta}}{\sqrt{[3]!(1+Q)^3}},
\end{equation}
where, $[n]!=[n][n-1]...[1]$. Notice the similarity of the numerators with the Hermite polynomials. In the limit $|f(n)|^2=1$ and $Q\longrightarrow1$, the recurrence relation

\begin{figure*}[ht]
    \centering
    \captionsetup{width=0.9\linewidth}
    \begin{subfigure}[t]{0.5\textwidth}
        \centering
        \includegraphics[height=2.2in]{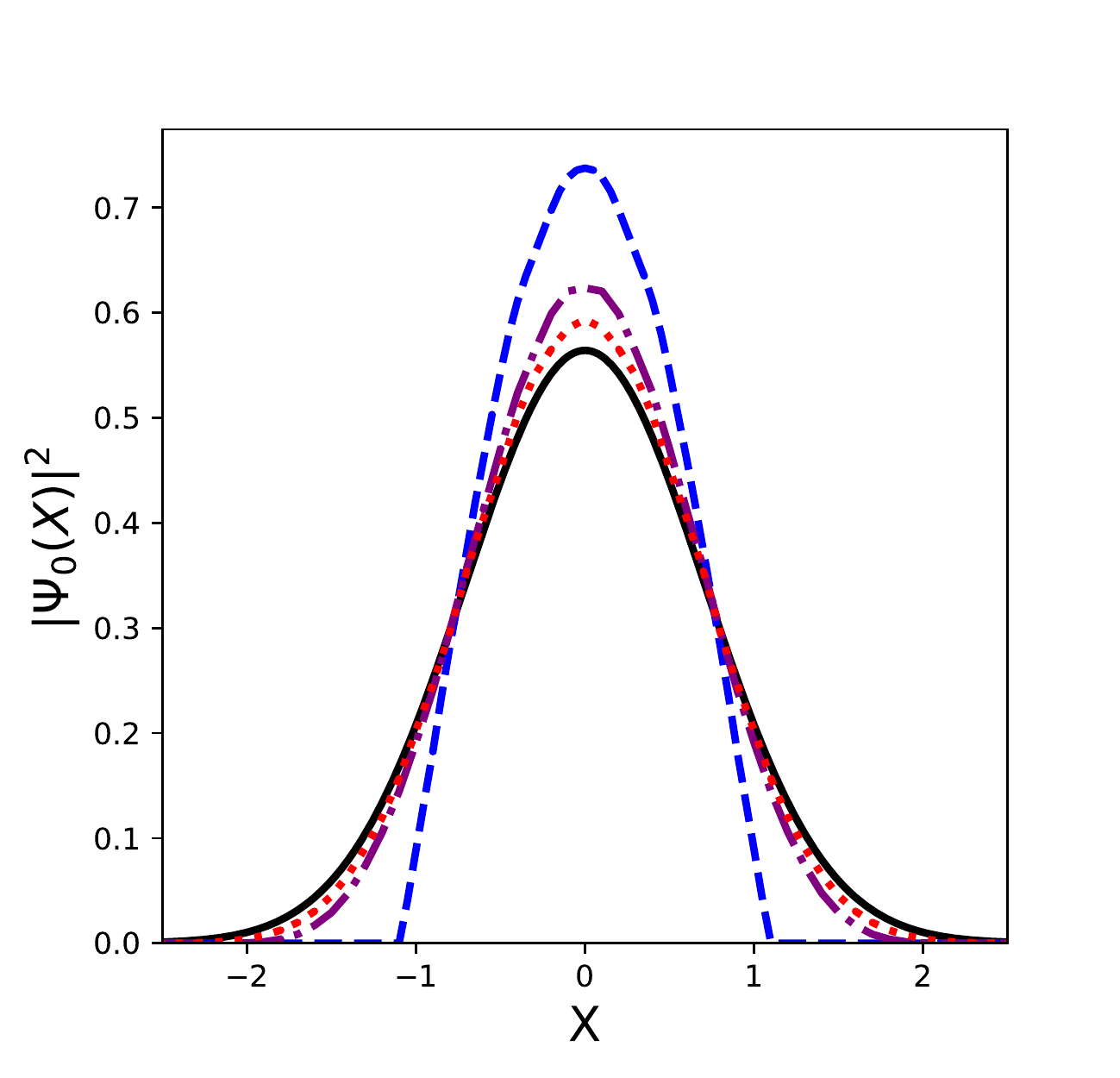}
        \caption{}
    \end{subfigure}%
    ~
    \begin{subfigure}[t]{0.5\textwidth}
        \centering
        \includegraphics[height=2.2in]{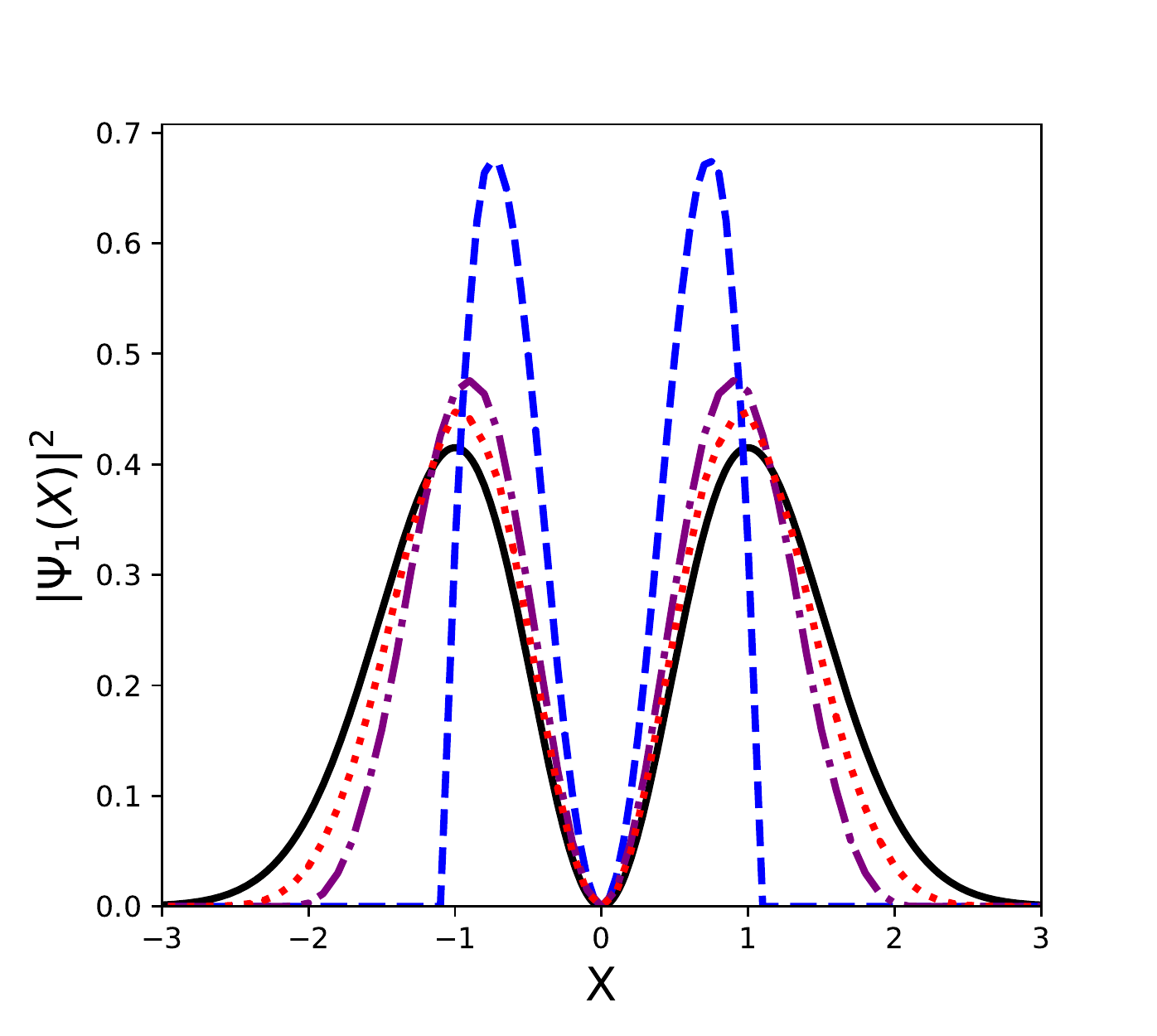}
        \caption{}
    \end{subfigure}
    \caption[6pt]{(a) ground state and (b) first excited state probability distribution functions for harmonic oscillator (solid), deformed oscillator with q=0.90 (dotted), q=0.80 (dashed-dotted) and q=0.30 (dashed) in the case of math type q-deformation.}
    \label{fig:mqdef}%
\end{figure*}

(\ref{J_nf}) reduces to the recurrence relation of the Hermite polynomial:
\begin{equation}
H_{n+1}(x_\theta)=2x_\theta \; H_n(x_\theta)-2n \;H_{n-1}(x_\theta).
\end{equation}
In the above limit, (\ref{Psi_nf}) also reduces to the quadrature representation of the non-deformed Fock state $|n\rangle$:
\begin{equation}
\Psi_{n}(X_\theta\rightarrow x_\theta)=\frac{H_n(x_\theta)}{\pi^{1/4} \; 2^{n/2}\sqrt{n!}}e^{-in\theta}e^{-x_\theta ^2/2},
\end{equation}
where $H_n(x_\theta)$ is the Hermite polynomial of order n. The ground state in the above limit is given by:
\begin{equation}
\Psi_{0}(X_\theta\rightarrow x_\theta)=\frac{e^{-x_\theta ^2/2}}{\pi^{1/4}}.
\end{equation}

Now, we can obtain an explicit representation for quadrature operator eigenstates  $|X_\theta\rangle$ in (\ref{fock_rep_xtheta}), in terms of the orthogonal polynomials $J_{n}(X_\theta)$:
\begin{equation}
|X_\theta\rangle=\sum\limits_{n=0}^{\infty}|n\rangle _{f} \; _{f}\langle n|X_\theta\rangle\\
=\overline{\Psi}_{0}(X_\theta)\sum\limits_{n=0}^{\infty}J_{n}(X_\theta) \; e^{in\theta} \; |n\rangle_f.
\end{equation}
Using the general expressions derived above, we obtain the polynomials $J_{n}(X_\theta)$ and wavefunctions $\Psi_{n}(X_\theta)$ for different types of deformations in the next section.

\subsection{Math-type $q$-deformed oscillator wavefunctions}

For the math-type $q$-deformation described by the deformed algebra (\ref{mathq_comm}), we had seen that $Q=q^2$ and $[n]=(1-q^{2n})/(1-q^2)$. Therefore in this case, the $q$-deformed Fock state in the quadrature basis becomes
\begin{equation}
\Psi_{n}(X_\theta)=e^{-in\theta} \; J_{n}(X_\theta) \; \Psi_{0}(X_\theta),
\end{equation}
where, the polynomial $J_{n}(X_\theta)$ has reduced to
\begin{equation}
J_{n+1}(X_\theta)=\frac{1}{\sqrt{[n+1]}}\bigg[\frac{2X_\theta}{\sqrt{1+q^2}}J_{n}(X_\theta)-\sqrt{[n]}J_{n-1}(X_\theta)\bigg].
\end{equation}
From the above equation, an explicit expression for  the deformed Fock state in the position representation
($\theta=0$) is given by
\begin{equation}
\Psi_{n}(X)=J_{n}(X)\Psi_{0}(X).
\end{equation}

In the limit $q \longrightarrow1$, $\Psi_{n}(X)$ reduces to the non-deformed harmonic oscillator position wavefunction:
\begin{equation}\label{Psi_nq_1}
\Psi_{n}(X\longrightarrow x)=\frac{H_n(x)}{\pi^{1/4}2^{n/2}\sqrt{n!}}e^{-x ^2/2},
\end{equation}
where, $H_n(x)$ is the Hermite polynomial of order n.

\par Figures \ref{fig:mqdef}.(a) and \ref{fig:mqdef}.(b) show the  deformed normalized ground state  and first excited state position probability densities, respectively for different deformation values. It can be seen that as $q$ value decreases (deformation increases), the peak of the probability curve increases in both figures. Moreover, in each case as $q\longrightarrow1$, the probability curve is seen to become the non-deformed harmonic oscillator probability distribution function.

\subsection{Physics-type $q$-deformed oscillator wavefunctions}
In the physics-type $q$-deformation described by the algebra (\ref{phyq_comm}) with $Q=q$ and $[n]={q(q^{-n}-q^{n})}/(1-q^2)$, the $q$-deformed Fock state in the quadrature basis is given by
\begin{equation}
\Psi_{n}(X_\theta)=e^{-in\theta} \; J_{n}(X_\theta) \; \Psi_{0}(X_\theta),
\end{equation}

\begin{figure*}[ht]
    \centering
    \captionsetup{width=0.9\linewidth}
    \begin{subfigure}[t]{0.5\textwidth}
        \centering
        \includegraphics[height=2.2in]{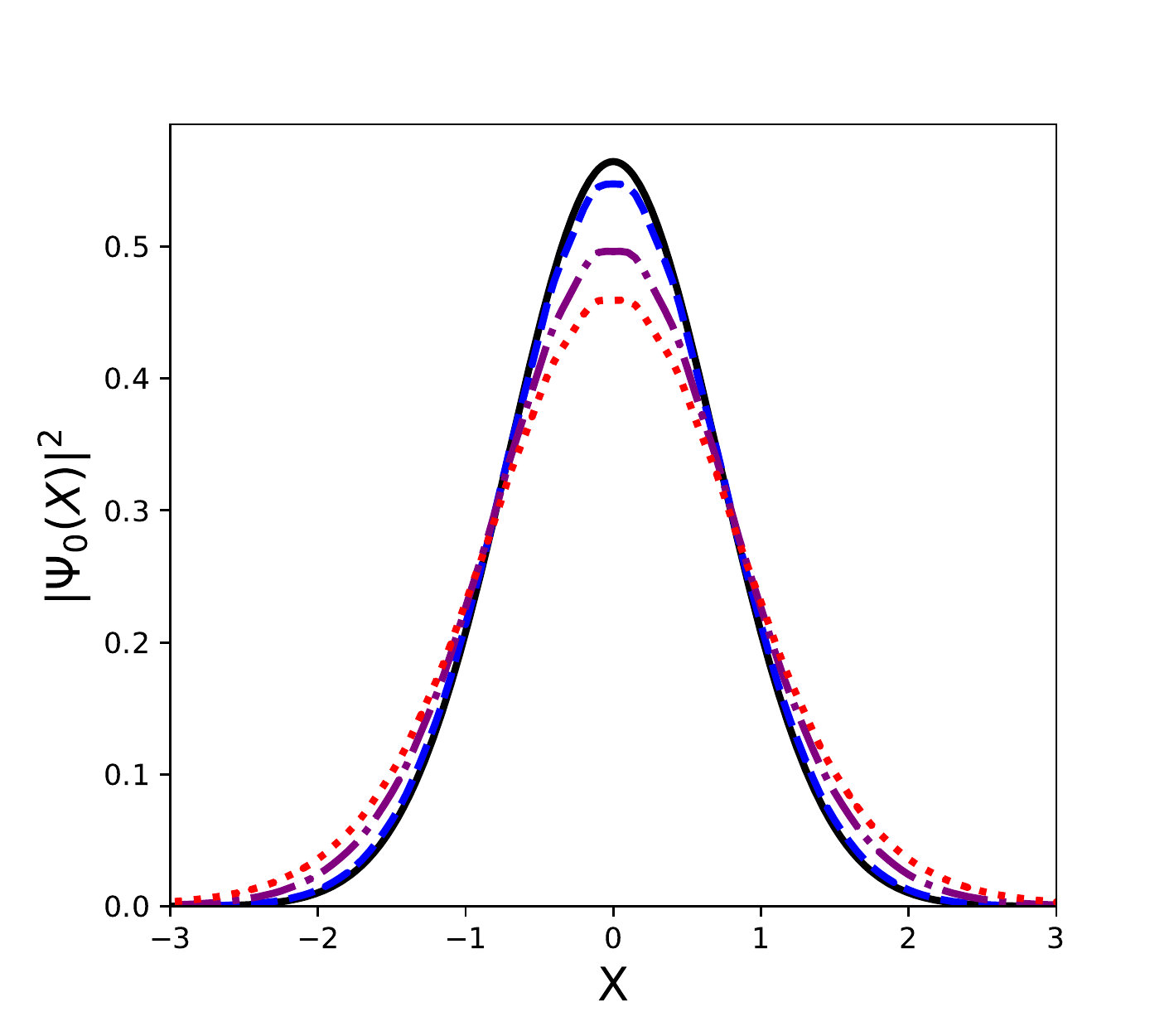}
        \caption{}
    \end{subfigure}%
    ~
    \begin{subfigure}[t]{0.5\textwidth}
        \centering
        \includegraphics[height=2.2in]{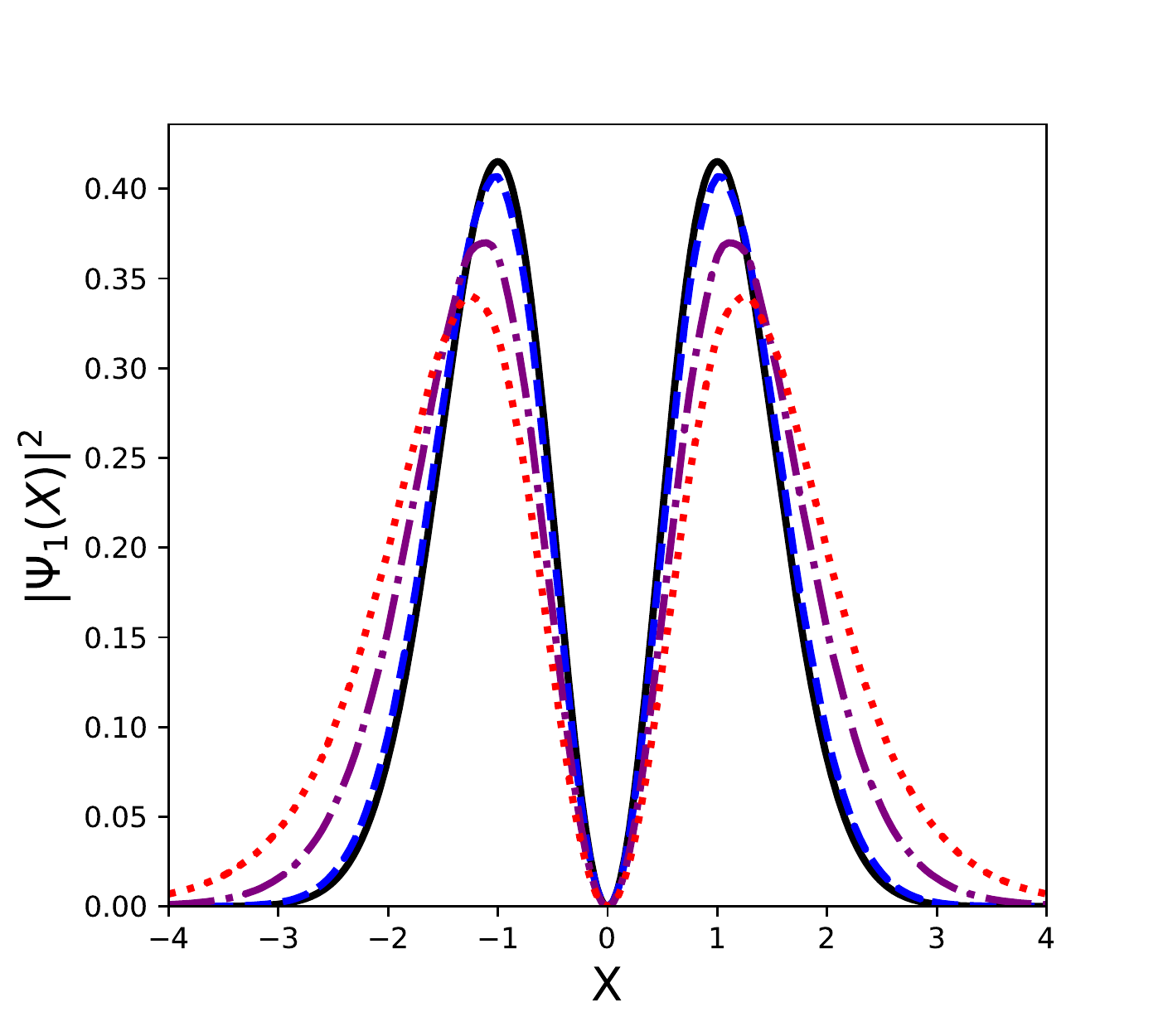}
        \caption{}
    \end{subfigure}
    \caption[4pt]{(a) ground state and (b) first excited state probability distribution functions for harmonic oscillator (solid), deformed oscillator with q=1.1 (dashed), q=1.5 (dashed-dotted) and q=1.9 (dotted) in the case of physics type q-deformation.}
    \label{fig:phqdef}%
\end{figure*}

where the polynomial $J_{n_f}(X_\theta)$ has reduced to
\begin{equation}
J_{n+1}(X_\theta)=\frac{1}{\sqrt{[n+1]}}\bigg[\frac{2X_\theta}{\sqrt{1+q}}J_{n}(X_\theta)-\sqrt{[n]}J_{n-1}(X_\theta)\bigg].
\end{equation}

In the position representation the deformed Fock state has the form,
\begin{equation}
\Psi_{n}(X)=J_{n}(X)\Psi_{0}(X),
\end{equation}
which reduces to (\ref{Psi_nq_1}) when $q$ $\rightarrow$ 1.

\par The plots of  the normalized position probability densities of the ground state  and first excited state  for different $q$ values  are shown in Figures \ref{fig:phqdef}.(a) and \ref{fig:phqdef}.(b), respectively. We find that the probability density for the harmonic oscillator displays the highest peak. In this case, as the $q$ value increases (deformation increases), the height of the peak is seen to decrease in contrast to the math-type $q$-deformation studied in the previous subsection. Moreover, the curve approaches the distribution function corresponding to the harmonic oscillator wavefunction as $q\longrightarrow1$ in both figures.

\subsection{($p$,$q$)-deformed oscillator wavefunctions}
In the case of ($p$,$q$)-deformation described by the algebra (\ref{pq_comm}) with $Q=p$ and $[n]=\frac{q(q^{-n}-p^{n})}{(1-pq)}$, the ($p$,$q$)-deformed Fock state in the quadrature representation is given by,
\begin{equation}
\Psi_{n}(X_\theta)=e^{-in\theta} \; J_{n}(X_\theta) \; \Psi_{0}(X_\theta),
\end{equation}
where, the polynomial $J_{n_f}(X_\theta)$ has reduced to
\\
\begin{equation}
J_{n+1}(X_\theta)=\frac{1}{\sqrt{[n+1]}}\bigg[\frac{2X_\theta}{\sqrt{1+p}}J_{n}(X_\theta)-\sqrt{[n]}J_{n-1}(X_\theta)\bigg].
\end{equation}

\par In the position representation, the deformed Fock state is given by,
\begin{equation}
\Psi_{n}(X)=J_{n}(X)\Psi_{0}(X),
\end{equation}
which reduces to (\ref{Psi_nq_1}) when ($p$,$q$) $\rightarrow$ 1.

\par
\begin{figure*}[h!]
    \centering
    \captionsetup{width=0.9\linewidth}
    \begin{subfigure}[t]{0.5\textwidth}
        \centering
        \includegraphics[height=2.2in]{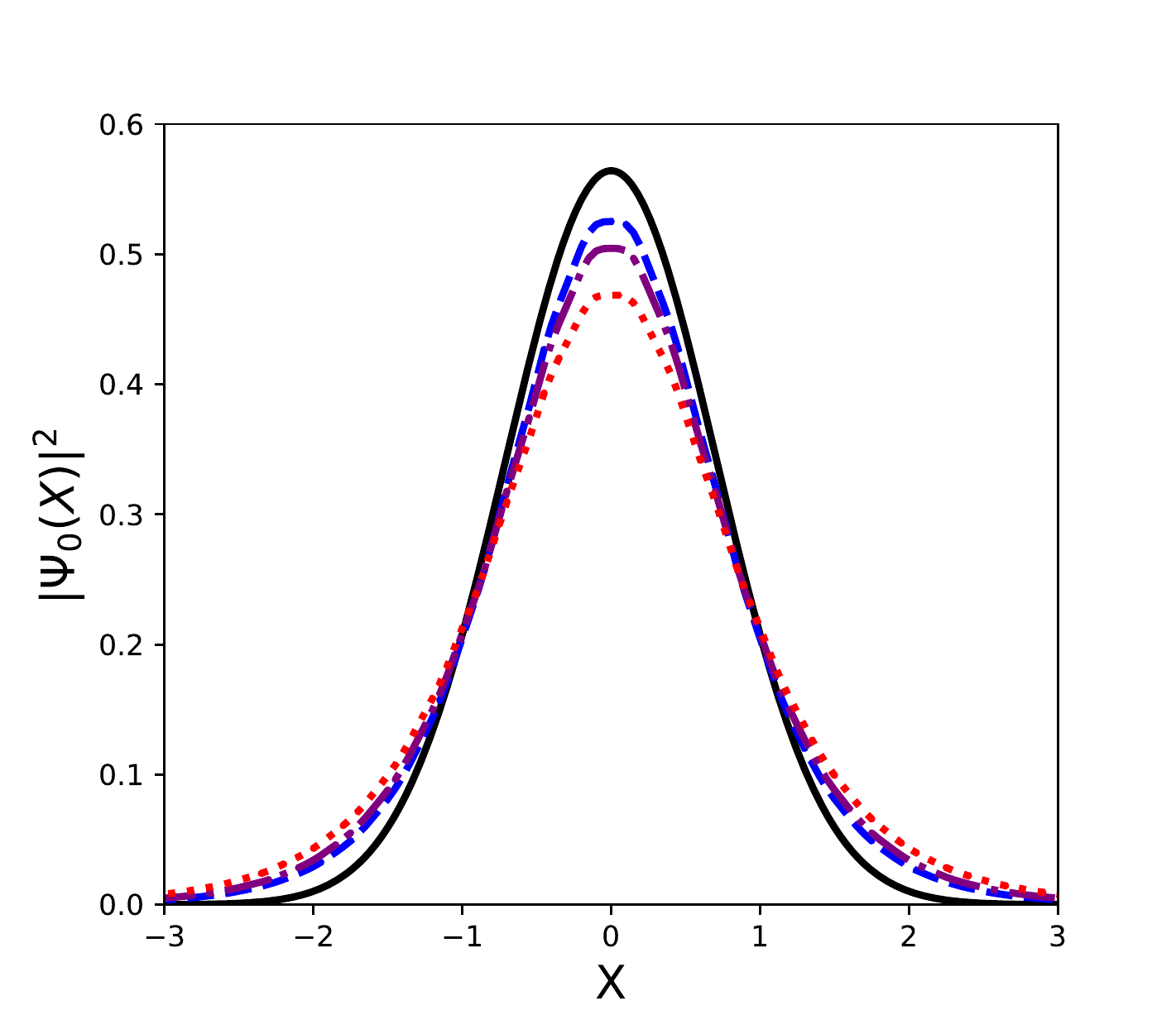}
        \caption{}
    \end{subfigure}%
    ~
    \begin{subfigure}[t]{0.5\textwidth}
        \centering
        \includegraphics[height=2.2in]{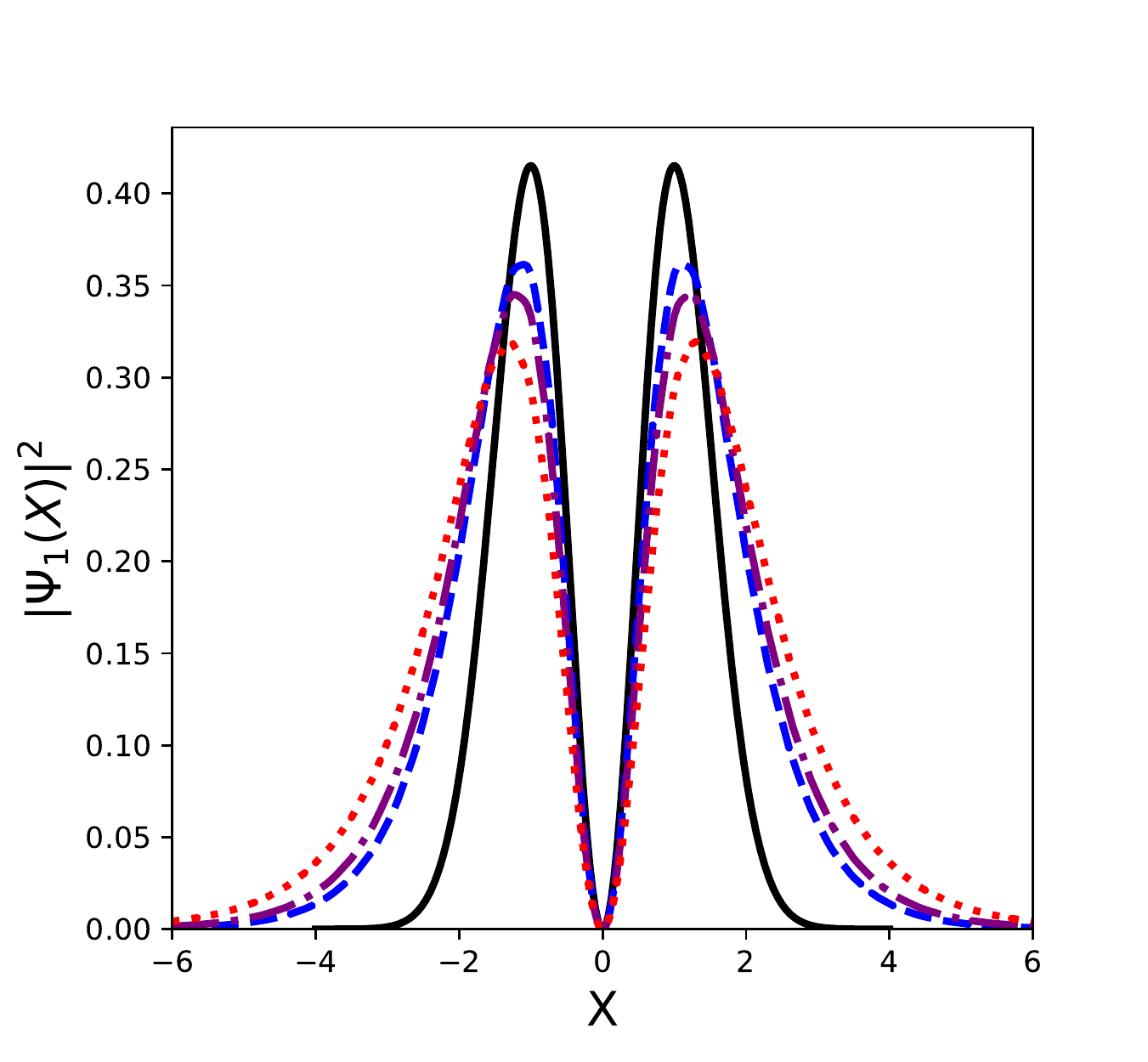}
        \caption{}
    \end{subfigure}
    \caption{(a) ground state and (b) first excited state probability distribution functions for harmonic oscillator(solid), deformed oscillator with p=1.3 q=0.5 (dashed), p=1.5 q=0.5 (dashed-dotted) and p=1.9 q=0.5 (dotted) in the case of (p,q)-deformation.}
    \label{fig:pqdef}%
\end{figure*}

 The normalized ground state and first excited state position probability densities for ($p$,$q$)-deformation are plotted for the same $q$ value and different $p$ values in Figures \ref{fig:pqdef}.(a) and \ref{fig:pqdef}.(b), respectively. We observe that for the same $q$ value, as $p$ increases (deformation increases), the height of the peak decreases. Further, the densities tend to the harmonic oscillator probability distribution when ($p$,$q$) $\rightarrow$ 1.

\section{Conclusion}
In this paper, we have provided a method to obtain the quadrature operator eigenstates  and wavefunctions for the general $f$-oscillators.
We have  defined the $f$-deformed quadrature operator with the help of the homodyne detection  method and represented its eigenstates in the $f$-deformed Fock state basis. This allowed us to produce a recurrence relation for the wavefunctions, which in turn allowed us to discover a new class of orthogonal polynomials $J_{n}(X_\theta)$. These new polynomials enabled us to represent the excited state wavefunctions of the $f$-oscillators in terms of the ground state wavefunction $\Psi_{0}(X_\theta)$. The polynomials $J_{n}(X_\theta)$ were found to be similar in properties of  the Hermite polynomials $H_{n}(x)$.

 We then considered different types of deformation and studied how the form of the polynomial $J_{n}(X_\theta)$ (and thus the form of the deformed wavefunction $\Psi(X_\theta)$) varies with the nature of deformation. Three types of deformed systems, namely math-type $q$-deformation, physics-type $q$-deformation and $(p,q)$-deformation,  were studied by appropriate substitution for $|f(n)|^2 $ and $Q$ in the expressions for the general $f$-oscillator. We also plotted the position probability distributions for the deformed ground state and the first excited state for each type of deformation that were studied. By comparing the probability curves for different values of the deformation parameter in each case, we demonstrated the difference between the deformed and the non-deformed harmonic oscillator wavefunctions. Our results will immensely help in the reconstruction of deformed quantum states  in experiments and quantum information theory based on deformed oscillators.

%%References section
%\begin{thebibliography}{99}

\end{document}